\documentclass[sigconf]{acmart}
\usepackage[utf8]{inputenc}
\usepackage{graphicx}

\title{A Serverless Architecture for Efficient and Scalable Monte Carlo Markov Chain Computation}

\author{Castagna Fabio}
\affiliation{
\department{Department of Pure and Applied Sciences, Computer Science Division}
\institution{Insubria University }
\postcode{21100}
\city{Varese}
\country{Italy}
}
\email{fcastagna@studenti.uninsubria.it}
\additionalaffiliation{
\institution{INAF–Osservatorio Astronomico di Brera}
\streetaddress{via Brera 28}
\postcode{20121}
\city{Milano}
\country{Italy}
}

\author{Trombetta Alberto}
\affiliation{
\department{Department of Pure and Applied Sciences, Computer Science Division}
\institution{Insubria University}
\postcode{21100}
\city{Varese}
\country{Italy}
}

\author{Landoni Marco}
\affiliation{
\institution{INAF–Osservatorio Astronomico di Brera}
\streetaddress{via Emilio Bianchi 46}
\postcode{23807}
\city{Merate}
\country{Italy}
}

\author{Andreon Stefano}
\affiliation{
\institution{INAF–Osservatorio Astronomico di Brera}
\streetaddress{via Brera 28}
\postcode{20121}
\city{Milano}
\country{Italy}
}


\begin{abstract}
Computer power is a constantly increasing demand in scientific data analyses, in particular when Markov Chain Monte Carlo (MCMC) methods are involved, for example for estimating integral functions or Bayesian posterior probabilities. In this paper, we describe the benefits of a parallel computation of MCMC using a cloud-based, serverless architecture: first, the computation time can be spread over thousands of processes, hence greatly reducing the time the user should wait to have its computation completed.
Second, the overhead time required for running in parallel several processes is minor and grows logarithmically with respect to the number of processes. Third, the serverless approach does not require time-consuming efforts for maintaining and updating the computing infrastructure when/if the number of walkers increases or for adapting the code to optimally use the infrastructure. The benefits are illustrated with the computation of the posterior probability distribution of a real astronomical analysis.
\end{abstract}

\copyrightyear{2023}
\acmYear{2023}
\setcopyright{rightsretained}
\acmConference[ICCBDC 2023]{2023 7th International Conference on Cloud and
Big Data Computing}{August 17--19, 2023}{Manchester, United Kingdom}
\acmBooktitle{2023 7th International Conference on Cloud and Big Data
Computing (ICCBDC 2023), August 17--19, 2023, Manchester, United
Kingdom}\acmDOI{10.1145/3616131.3616141}
\acmISBN{979-8-4007-0733-9/23/08}

\begin{document}

\newcommand{\ppf}{\texttt{PreProFit}}
\maketitle

\section{Introduction}
Computing power is a resource in ever-increasing demand given the relevance of data analysis and simulations in scientific settings.
In order to provide the desired amount of computing power, in the past decades several large, ad-hoc scientific computing infrastructures have been devised, including HPC centers, cluster and/or grids. On a smaller scale, on-premises approaches have been applied for scientific computing facilities at universities and research centers.
In commercial and industrial settings, the cloud computing paradigm is the primary mean of accessing distributed computing resources \cite{tangirala2016}, as leveraged from large-scale datacenters.
In the past years, the commercial offering of cloud-based computing resources has evolved from virtualization and storage services to containerization and orchestration of cloud-native services commonly known as "serverless computing". In a serverless setting, functions are executed with minimal server-side overhead. This allows for their combination in  "microservice-based architectures". Whereas we acknowledge that the serverless computing model has not been designed to support scientific computing (since the workloads executed in a cloud-native way are typically lightweight), there are several advantages in deploying this computing model in scientific scenarios. Among the most relevant ones:
functions provide the right abstraction level for problems arising in scientific contexts; the burden of managing the complex infrastructure necessary for ensuring the serverless approach is abstracted away from the application programming level and -- most importantly -- resources are automatically managed in a highly elastic fashion. The last point is particularly relevant for serving a large amount of fine-grained requests
in a responsive way, as opposed to more traditional ways of providing distributed computing resources, as Infrastructure as a Service (IaaS) or High Performance Computing (HPC).
Of course, the serverless approach has its own drawbacks, mainly due to platform lock-in (since applications using serverless functions rely on the APIs provided by the vendors) and scarce control over the platform itself.

Monte Carlo Markov Chain (MCMC) \cite{Hastings1970} is perhaps the most common numerical algorithm to sample a function, often a Bayesian posterior probability or a function to be integrated. When converged to the stationary distribution, an MCMC returns samples with a density proportional to the value of the target function and with accuracy increasing with sample size. The MCMC achieves its purpose by constructing a Markov chain, i.e. 
a stochastic process in which the acceptance of a new sample only depends on the target function values 
at a newly proposed sample and at the previous one.
Finally, all the chains are usually merged together, hence improving the closeness to the target function.
In its entirety, this procedure encompasses two components: one that is definitely sequential (each iteration depends on the previous one), and one that is potentially parallelizable: different chains (or \emph{walkers}) may be run independently from each other. The crux of the problem stands in finding a good compromise when selecting the number of iterations and walkers of the MCMC, while keeping the convergence under control \cite{Jacob2010}.
In the past years, there have been many attempts to obtain more efficient MCMC computations, mostly relying on the opportunity to parallelize different walkers \cite{vanderwerken2013}, e.g. via importance sampling or rejection sampling. Nonetheless, this approach has the drawback of preventing the chains from communicating with each other. Also, the required effort from the programmer to adapt the computing infrastructure along with the corresponding code as the demand for parallelization scales up, is not trivial.

In this work we present a novel serverless architecture tailored for the efficient, parallel computation of MCMC using a cloud-based, serverless architecture. As it will be explained in the following, the benefits of this approach are: (i) the overhead required for running in parallel several walkers grows logarithmically with respect to the number of walkers; (ii) the serverless approach does not require time-consuming efforts for maintaining and updating the computing infrastructure when/if the number of walkers increases or for adapting the code to optimally use the infrastructure.

The paper is organized as follows: in Section \ref{sec:relworks} a short overview of the relevant works is presented; in Section \ref{sec:archi} we present a serverless architecture apt for MCMC computations; in Section \ref{sec:exp} we present the experimental setting over which the architecture has been tested and we discuss the obtained results; finally in Section \ref{sec:conc} we draw our conclusion and we point to some future work directions.

\section{Related works}
\label{sec:relworks}
In recent years, researchers across the scientific community have started to investigate the feasibility of adopting the serverless computing approach in data analytics applications.
Jonas et al. \cite{DBLP:conf/cloud/JonasPVSR17} presented \texttt{PyWren}, a prototype system that leverages stateless functions for embarrassingly parallel computations and makes use of remote storage.
\texttt{NumPyWren} \cite{DBLP:conf/cloud/ShankarKVPRSRJV20} emerged as an extension of the previous framework, meant to perform large-scale linear algebra operations.
The execution of stateless functions through cloud computing models, the so-called function-as-a-service paradigm, has been applied also for scientific workflows: Malawski et al. \cite{DBLP:journals/fgcs/MalawskiGZBF20} experimented the use of \texttt{HyperFlow}, a lightweight workflow engine, for such purposes, while Arjona et al. \cite{DBLP:journals/fgcs/ArjonaLSSV21} carried out similar work with \texttt{TriggerFlow}, a trigger-based event-driven workflow framework.
An interesting project of using serverless infrastructures as back-end to domain-specific scientific tool has been developed at CERN: \texttt{ROOT Lambda} \cite{DBLP:conf/ccgrid/KusnierzPMBSAPA22} is an engine built upon ROOT, a specific software for analyzing High-Energy Physics data, and relies on AWS Lambda platform for serverless computing.
Another considerable example is \texttt{FuncX} \cite{DBLP:journals/tpds/LiCBGSNWBBKFC22}, whose great potential is the opportunity to execute the functions across a federated ecosystem of endpoints.
Other works have emerged with the aim to implement the MapReduce paradigm within a serverless architecture: \texttt{Flint} \cite{Kim2018}, \texttt{Qubole}\footnote{\url{https://github.com/qubole/}}, \texttt{Corral}\footnote{\url{https://github.com/bcongdon/corral}}, \texttt{Lambada}\footnote{\url{https://gitlab.com/josefspillner/lambada}}, \texttt{Apache Pulsar}\footnote{\url{https://pulsar.apache.org/docs/2.11.x/}}, \texttt{Crucial} \cite{barcelona2019} are some examples worth to be mentioned.

More recently, it has been proposed that those projects, and others similar to them, should be actually viewed as hybrid systems that combine both serverless and non-serverless components and introduced the \textit{ServerMix} model, that integrates them in a single framework \cite{Garcia-Lopez2019}. 
Another interesting proposal is \texttt{Lithops} \cite{Sampe2021}, within the EU CloudButton project\footnote{\url{https://cloudbutton.eu/}}. 
Lithops is an advanced, single, unified framework that is able to scale massive single-machine code computations to cloud platform executors. \texttt{Lithops} relies on two independent cloud components, a computational engine back-end and a storage back-end.

The bottom line is that currently there is no consensus on a single, comprehensive approach that mediates between serverless and non-serverless approaches, taking care of the trade-offs in terms of disaggregation, isolation, and scheduling of the resources. It is hinted though that in the near future the trade-off will be progressively unbalanced in favor of a purely serverless approach.

\section{Serverless MCMC Architecture}
\label{sec:archi}
The key point that triggered our serverless MCMC approach is the observation that the computation of the posterior function in a generic MCMC process requires a substantial amount of computation, e.g. in the order of few tens of minutes, as in the case presented in Section \ref{sec:exp}. Almost its entire computational cost comes from the \emph{likelihood function}, which evaluates the fit of the parametrized model on the observed data. 
Within this time frame, any attempt to parallelize the MCMC process would be dominated by the complexity of the likelihood function itself, which is required to be computed a few hundred thousand times in order to get a precise estimate of the posterior distribution. Usually, one could attempt to parallelize the MCMC computation by using an \emph{embarrassingly parallel} technique \cite{vanderwerken2013}
in which each walker is independently executed on each CPU core. Although this approach is quite simple, it has the drawback that its scalability towards hundreds of walkers is not trivial, especially when running the code on distributed environments such as HPC. It is necessary to carefully assess the architecture of the cluster (number of cores, nodes, shared file systems, etc.) and communication between walkers is hard to be achieved in a clean way (shared files, etc.).  Moreover, when the likelihood function is hard to compute, the trade-off between the number of walkers and the number of iterations of each walker becomes critical and these should be balanced for reaching convergence in a reasonable amount of time.

Thus, if one could compute in parallel an arbitrary number of likelihood functions, with reduced overheads, without having the need to a-priori knowing how many of them shall be computed, an MCMC computation could be parallelizable and scalable with communication between walkers all at the same time in a completely serverless fashion. 

\begin{figure*} [ht!]
\begin{center}
\begin{tabular}{c}
\includegraphics[scale=0.84]{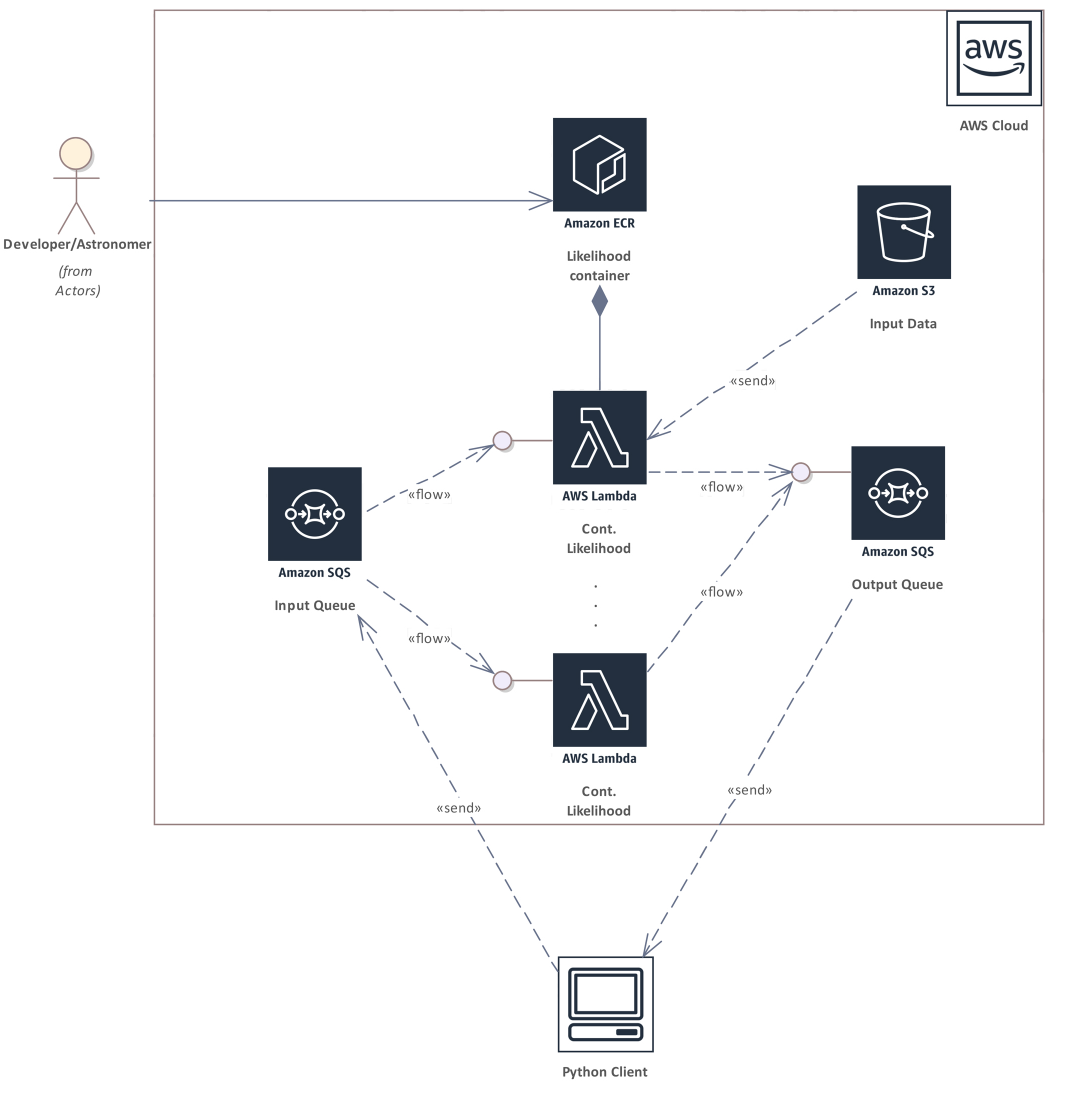}
\end{tabular}
\end{center}
\caption{The Serverless architecture for the MCMC on AWS. In this scenario, the computationally hard likelihood function at the core of the MCMC is loaded as a Docker Container in Amazon ECR by the programmer. A simple Python application that implements the less computationally intensive part of the MCMC pushes into the Input SQS queue the necessary information for computing an arbitrarily large number of likelihood functions in parallel (proportional to the number of walkers in the MCMC). A trigger configured on Input SQS queue instantiates an Amazon Lambda Function for each input message that is based on the container uploaded in Amazon ECR. The output of each AWS Lambda function is stored in the output queue and fetched by the Python application. Input data required to compute the likelihood (e.g. astronomical input images) are shared between the various AWS Lambda functions using a bucket in Amazon S3.} {\label{fig:framework}}
\end{figure*} 

\begin{figure*}[ht!]
\begin{center}
\begin{tabular}{c}
\includegraphics[width=10 truecm]{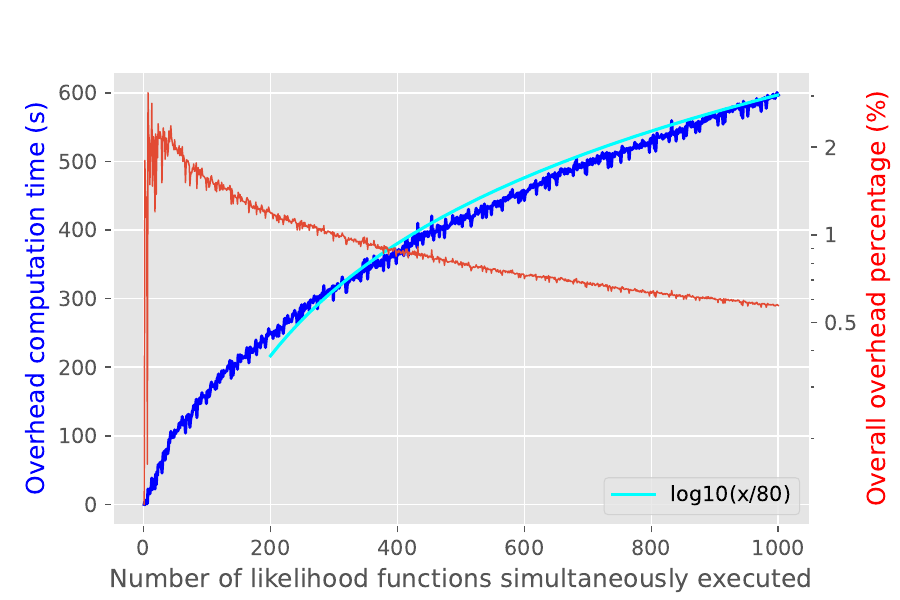}
\end{tabular}
\caption{Scalability plot. Blue curve: overhead computation time defined as the interval between the arrival time of the first and the last likelihood outputs.
Red curve: overhead percentage on the overall system, obtained as the ratio between the blue curve and the typical execution time, 100s.
The two curves show some jittering because of minor differences across the independent runs (one per number of likelihood simultaneously executed).} \label{fig:time_lambdas}
\end{center}
\end{figure*} 

Following this line of reasoning, we implemented a serverless architecture (Fig.~\ref{fig:framework}), Amazon Web Services (AWS) cloud computing infrastructure. It is important to stress here that we do not rely on ad-hoc features of AWS and our approach is easily reproducible on any cloud computing infrastructure like Google Cloud or Microsoft Azure, as well as on equivalent open-source solutions (e.g., Apache OpenWhisk\footnote{\url{https://openwhisk.apache.org/}}, Supabase\footnote{\url{https://supabase.com/}}, MinIO\footnote{\url{https://min.io/}}, Knative\footnote{\url{https://knative.dev/docs/}}).
In detail, we use Docker to containerize the Python-based code of our likelihood function, which is described in Section \ref{sec:exp}.
The likelihood function receives its input parameter and writes the output by means of queues as explained below. We make use of Amazon Container Registry (ECR) in Fig.~\ref{fig:framework} for storing the Docker image (and its relative version).
The execution of an MCMC in our Serverless architecture is explained in rather simple terms: a Python application, that maintains the status of each walker (the number of walkers can be arbitrarily large), pushes the input into the Amazon Simple Queue Service (SQS, FIFO) Input Queue as a message that contains the relevant input needed by a likelihood to obtain the result. When a message in the queue is pushed, a trigger instantiates a lambda function based on the container in which the source code of the likelihood function is stored and the computation is automatically started. The output of each likelihood at the end of the process is stored as a message in the SQS output queue and delivered to the Python application that -- since it maintains the status of each walker -- has a global view of all of the walkers. This entails the possibility of communication between walkers only at the expense of the small communication overhead between the queues. 
The input raw data required by each likelihood are stored in Amazon Amazon Simple Storage Service (S3), a highly scalable service perfectly suitable for the unprecedented amounts of data provided by new-generation astronomical surveys.

\section{Experimental setting and results}
\label{sec:exp}
We tested our approach using an array of cloud-based services offered by Amazon (as already mentioned in Section \ref{sec:archi}). Namely, Amazon Lambda, SQS, ECR and S3 on a real-world astrophysical dataset. We chose to work with astrophysical data.
As it happened for the majority of scientific disciplines, the availability of vast amounts of heterogeneous - often rapidly evolving - data has strongly affected the astronomical and astrophysical community. The number of instruments that allow the study of the Universe has progressively increased across the decades, and most importantly, the amount of data that they are able to collect has literally exploded in recent times. Simultaneously, such a huge data availability allows astronomers to investigate more complex questions, which in turn requires advanced analysis methods \cite{Feigelson2021}.

\begin{figure*}[ht!]
\begin{center}
\begin{tabular}{c}
\includegraphics[width=14 truecm]{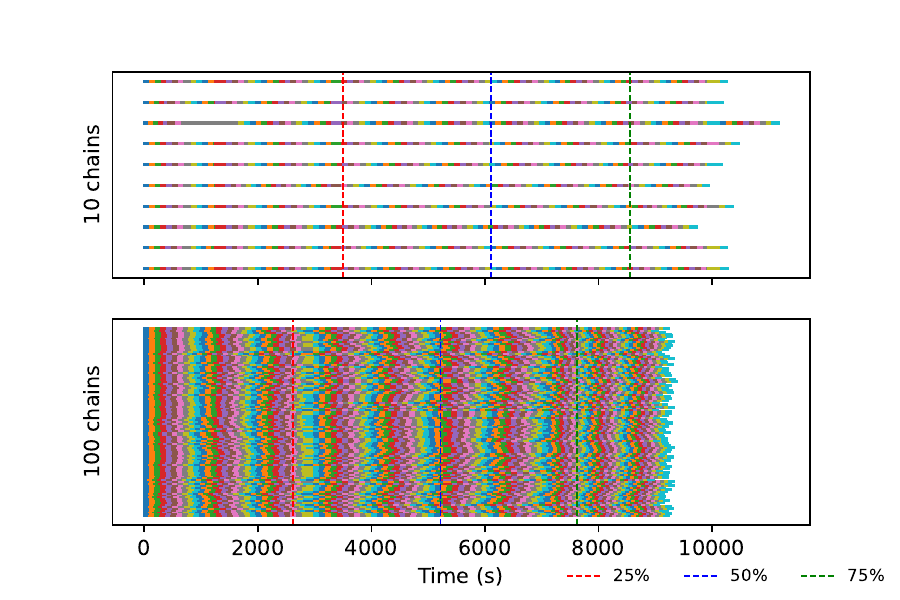}
\end{tabular}
\caption{Total waiting time to run multiple chains each with 100 iterations. The three vertical dashed lines show the intermediate waiting time after 25, 50, and 75\% of the iterations. Color bands are almost vertical, indicating that there is little overhead to moving to larger number of walkers.} \label{fig:timeline}
\end{center}
\end{figure*} 

We tested the proposed architecture with galaxy clusters data, taking advantage of \ppf\footnote{\url{https://github.com/fcastagna/preprofit}}\cite{Castagna2019}, a pre-existing tool that we adapted and included in our framework.
\ppf\, is a publicly available Python code designed to fit the pressure profile of a galaxy cluster through an MCMC algorithm. 
The likelihood function published in \cite{Castagna2019} performs the following tasks for one cluster:
(i) computes a polynomial function based on the input parameters,
(ii) applies a forward Abel transform (computationally equivalent to a one-dimensional integration), (iii) extends the problem from one-dimensional to two-dimensional setting, (iv) applies a convolution through Fast Fourier Transform, (v) performs some further minor operations and finally returns the overall likelihood value according to the Chi-Square evaluation.
Since the initial release of \ppf, its code has been widely updated and optimized, and - most importantly - has been extended from single-object analysis at a time to multiple galaxy clusters at the same time, meaning that parameter estimation is simultaneously conducted at both individual level and population of clusters level through a Bayesian hierarchical model. 
By increasing the number of analyzed objects, the overall computational effort for obtaining the requested likelihood justifies the deployment of a cloud-based, serverless architecture.
With about 200 galaxy clusters, the typical likelihood computation takes about 100 seconds.
In the near future, the demand will be even more pressing, given that existing surveys already provide observations of about one thousand galaxy clusters, and incoming ones will largely exceed in size.

In order to assess the performance of the proposed serverless MCMC solution, we first estimate the end-to-end overhead by measuring the required amount of time for computing, in parallel, a given number of likelihoods (keeping in mind that each requires roughly 100 seconds to be computed). The resulting time encompasses the duration of all the operations performed, from queuing the messages in the input queue to fetching the result from the output queue.
We used private datasets from 200 galaxy clusters observed by the South Pole Telescope \cite{Carlstrom2011} for approximately 200 MB, which include observed data and all other objects required for the likelihood computation as described in \cite{Castagna2019}. We remind here how the heavy amount of data at play does not reside in the raw data, but in the intermediate objects required within the likelihood computation.
The left axis of Figure \ref{fig:time_lambdas} shows the total overhead time as a function of the number of parallel likelihood functions. 
The total overhead time, defined as the interval between the arrival time of the first and the last likelihood outputs, grows in nearly logarithmic fashion when the number of computed likelihoods is larger than $\sim$ 200-300. The overhead time for computing in parallel several thousand likelihoods
is of the order of few hundreds of seconds (see Figure \ref{fig:time_lambdas}). The key point is that such overhead is a very small fraction of the overall time required to compute the whole chain.
Using 1000 walkers, the ratio between overhead time and overall time is less than 0.6\% (Fig.~\ref{fig:time_lambdas}, red line). This has relevant implications since, adopting the proposed MCMC serverless architecture, a user may aggressively parallelize the computation of an MCMC, using a large amount of walkers, without incurring into a significant amount of overhead. 
It is important to remark that the parallelization is performed without being bothered by maintaining the underlying computing infrastructure or by tailoring the (fitting) code to optimally exploit the available infrastructure.

We also report the results of various tests on complete MCMC executions, introducing the sequential component of the algorithm.
Figure~\ref{fig:timeline} shows the overall waiting time needed to run 100 iterations of each chain, with a varying number of chains, i.e. the amount of parallel likelihood executions at every step of the MCMC. We considered two different scenarios of 10 and 100 chains (equivalent to an overall number of 1000 and 10000 samplings, respectively).
The overall waiting time is approximately constant and comparable in both scenarios, despite the increasing number of likelihood executions (in the presented case the larger computation is even quicker than the smallest one). Furthermore, later started walkers are not delayed compared to earlier started walkers because the color bands in the bottom panel of  Fig.~\ref{fig:timeline} are almost vertical, suggesting that when increasing to, e.g., 1000 walkers, additional overhead
time will be minor at most.
This result strongly suggests that the overall execution times of entire MCMC computations stay almost constant as we add chains.

We are currently considering specific MCMC samplers that take advantage of the communication among walkers to improve the convergence of the chains and their closeness to the target function such as the block independent Metropolis-Hastings algorithm \cite{Jacob2010}. The idea is to exploit the parallelism of the architecture by running walkers independently, but also from time to time to exchange information among chains on the state of each random walker. This is efficient
in our setting because at each iteration the output of each simultaneous likelihood estimate is collected by the Python application.

\section{Conclusions}
\label{sec:conc}
In this work we have presented a serverless architecture for the parallel computation of MCMC. Our approach shows the feasibility of MCMC computations in a massively parallel fashion with very small overhead, without having to think about how to modify/improve the computing infrastructure as the computation scales to thousands of walkers executed in parallel.  
As future steps, we plan (i) to port our architecture from AWS-based to other cloud platforms (e.g. Apache OpenWhisk); (ii) to expand the analysis on larger datasets as they become available; (iii) to further investigate the communication across walkers at key points along the MCMC iterations, as this is relevant for improving the convergence of the chains and their closeness to the target function.


\bibliography{acmart}
\bibliographystyle{unsrtnat}%

\end{document}